\documentclass[12pt,twoside]{article}   
\usepackage[super,sort,comma]{natbib}

\usepackage{amsmath,amsfonts,amssymb}
\usepackage[pdftex]{graphicx}  
\usepackage{caption}
\usepackage{rotating}
\usepackage{tabularx}
\usepackage[colorlinks=true,linkcolor=blue,citecolor=blue]{hyperref}
\graphicspath{{figures/}}
\usepackage{array,multirow,graphicx}
\usepackage[inline,shortlabels]{enumitem} 
\usepackage{xcolor,soul}

\usepackage{tikz}
\usetikzlibrary{arrows,intersections}
\usetikzlibrary{math} 

\usepackage{fancyhdr}		

\usepackage[section]{placeins}   %

\makeatletter \renewcommand\@biblabel[1]{$^{#1}$} \makeatother
\newcommand{\cen}[1]{\begin{center} #1 \end{center}}

\usepackage[mathlines]{lineno}

\usepackage{hyperref}
\hypersetup{ colorlinks,
    citecolor=blue,
    filecolor=blue,
    linkcolor=blue,
    urlcolor=blue
}

\usepackage{xcolor}

\definecolor{gray}{rgb}{0.6,0.6,0.6}
\definecolor{red}{rgb}{0.85,0,0}
\definecolor{green}{rgb}{0,0.85,0}
\definecolor{blue}{rgb}{0,0,0.85}
\definecolor{beige}{rgb}{0.92,0.87,0.78}
\usepackage[all]{hypcap}    

\usepackage{subfig}
\usepackage{floatrow}
\begin{document}

\cen{\sf {\Large {\bfseries Improving the modeling of the Agility multi-leaf collimator}}} 
\vspace*{10mm}
\begin{flushleft}

Mohammad Hussein$^{1}$\\
Metrology for Medical Physics Center, National Physical Laboratory, Teddington, UK

\vspace*{3mm}

Agnes Angerud\\
RaySearch Laboratories AB, Stockholm, Sweden

\vspace*{3mm}
Jordi Saez \\
Department of Radiation Oncology, Hospital Cl\'inic de Barcelona, Barcelona, Spain \\

\vspace*{3mm}
Evelien Bogaert\\
Department of Radiation Oncology, Ghent University Hospital and Ghent University, Ghent, Belgium

\vspace*{3mm}
Matthieu Lemire\\
Department of Medical Physics, CIUSSS de l'Est-de-l'Île-de-Montr\'eal, Canada

\vspace*{3mm}
Miriam Barry\\
Metrology for Medical Physics Center, National Physical Laboratory, Teddington, UK

\vspace*{3mm}
Ileana Silvestre Patallo\\
Metrology for Medical Physics Center, National Physical Laboratory, Teddington, UK

\vspace*{3mm}
David Shipley\\
Metrology for Medical Physics Center, National Physical Laboratory, Teddington, UK

\vspace*{3mm}
Catharine H. Clark\\
Metrology for Medical Physics Center, National Physical Laboratory, Teddington, UK \\
Medical Physics, University College London Hospital, London, UK \\
Medical Physics and Bioengineering, University College London, London, UK

\vspace*{3mm}
Victor Hernandez\\
Department of Medical Physics, Hospital Sant Joan de Reus, IISPV, Tarragona, Spain \\
Universitat Rovira i Virgili (URV), Tarragona, Spain \\

\vspace*{5mm}
$^{1}$ Author to whom correspondence should be addressed. \\
email: mohammad.hussein@npl.co.uk

\end{flushleft}

Version typeset \today\\

\pagenumbering{roman}
\setcounter{page}{1}
\pagestyle{plain}


\newpage     
\begin{abstract}
\begin{flushleft}
\noindent {\bf Background:} Robust fine tuning of MLC TPS modeling parameters is crucial for creating an optimal beam model, particularly with the ever-increasing accuracy required for advancing techniques. Challenges can arise from balancing the trade-off between multiple parameters and therefore the quality of the parameter tuning will often depend on the experience of the physicist and on the procedures used. This is in part due to limitations of the MLC modeling within the TPS. As a result, the actual MLC parameter values used have been shown to vary widely between centers. A novel methodology which standardizes and simplifies the process of robust convergence to an optimized MLC model has previously been proposed. It is based on measurements of a set of dynamic fields with a Farmer-type chamber.\\
\noindent {\bf Purpose:} In this study we present and evaluate two different MLC transmission maps to improve the modeling of the Elekta Agility MLC in the RayStation TPS. This study investigates the impact of these improvements using test fields and a wide variety of measured clinical plans.\\
{\bf Materials:} Three MLC models were assessed: the current clinically available MLC model and two prototype MLC models with discrete (Prototype I) and continuous (Prototype II) transmission maps assigned to the tongue-and-groove and leaf tip regions. Both prototypes aimed to replicate the average doses for synchronous and asynchronous sweeping gap fields, measured using a Farmer chamber in a solid water phantom, at three different centers. Each center performed QA measurements of clinically relevant plans using their own device and software analysis techniques. The plans were calculated using each center's implementation of the current MLC model and using each prototype utilizing a common set of MLC parameters.\\
{\bf Results:} All MLC models achieved good accuracy in clinical plans using 6\,MV beams, with average gamma passing rates at 2\%/2\,mm of 96.1\% (clinical models), 97.0\% (Prototype I), and 97.0\% (Prototype II) across all centers, including a great variety of treatment plans and measuring systems. Both prototypes improved the agreement with measurements in some cases, notably for the synchronous and asynchronous sweeping gaps and for the  test. Additionally, the prototypes were easier to configure because no trade-offs were required, and only a slight machine-specific adjustment of the leaf offset parameter was needed.\\
{\bf Conclusions:} The two MLC prototypes created within RayStation for the Agility facilitated the standardization of the configuration and commissioning processes and extended the range of validity of TPS dose calculations. The simple MLC prototype with discrete transmission maps performed similarly to the more sophisticated one both in tests and clinical plans and constitutes a good option for routine clinical practice. Both prototypes reduced the need for trade-offs and were successfully configured using a common set of MLC parameters across different centers, which is useful for reducing the workload and the risks associated with the MLC configuration process, thus improving the accuracy and safety of radiotherapy treatments.\\
\end{flushleft}

\end{abstract}

\newpage     

The table of contents is for drafting and refereeing purposes only.
\tableofcontents

\newpage

\setlength{\baselineskip}{0.7cm}      

\pagenumbering{arabic}
\setcounter{page}{1}
\pagestyle{fancy}

\section{Introduction}
\begin{flushleft}
Robust fine-tuning of multi-leaf collimator (MLC) modeling parameters in radiotherapy Treatment Planning Systems (TPS) is crucial for creating an optimal IMRT and/or VMAT beam model, particularly with the ever-increasing accuracy required for advancing techniques.

Challenges can arise from limitations in the MLC model implemented in the TPS and from the need to balance trade-offs between multiple TPS parameters. Additionally, there is a lack of standardization of the procedures used for the configuration and commissioning of MLC models within TPSs. There is a wide variety of approaches from using simple test fields that aim to characterize key features of the MLC such as the dosimetric leaf gap (DLG), transmission, leaf offset etc. to using clinically relevant plans which are then used to iteratively fine-tune some set of modeling parameters. Limitations in MLC modeling can make it necessary to deviate from physically realistic parameters as measured by simple static tests to compensate for calculation differences when modulation is introduced. As a result, the quality of the parameter tuning will often depend on the experience of the physicist and the type of tests carried out, which explains why the actual MLC parameter values used in the clinic vary widely between centers, which can lead to dose calculation errors\citep{Kerns:2017aa,Glenn:2020aa,Glenn:2022aa}.

A novel methodology for robust convergence to an optimized MLC model has previously been proposed based on measurements of a set of dynamic fields with a Farmer-type chamber\cite{Saez:2020aa,Hernandez:2017aa}. The method has been shown to be able to accurately characterize fine details of the MLC such as tongue-and-groove width and leaf tip width. Using this methodology, Saez et al. described how to extract RayStation parameters to reproduce the average doses of synchronous and asynchronous sweeping gaps for Varian's Millennium and HD120 MLCs\cite{Saez:2020aa}. 

Where the Varian TrueBeam and the STx MLC both have a stepped tongue-and-groove profile design, the Agility MLC uses a flat slightly non-target pointing surface\cite{Yu:1998aa}. This tilted leaf side is cut at the top and the bottom by the rounding of the leaf tip, reducing both the thickness at the center of the leaf and the maximum extension of the tongue-and-groove into the open region\cite{Hernandez:2022aa, Thompson:2014aa, Gholampourkashi:2019aa}. It was recently reported that these characteristics produce dosimetric effects associated with the rounded leaf end and progressively increasing tongue-and-groove shadowing as far as 20\,mm from the leaf tip end\cite{Hernandez:2022aa}. The RayStation and Monaco TPSs did not account for this effect, and in neither system is there one set of parameters that provides a full characterization of the Agility. Consequently, trade-offs in the MLC parameters for the Agility MLC are necessary, which also contributes to the great variability in the final MLC parameters used by different centers\cite{Hernandez:2022aa}.

In this study we present and evaluate two different MLC transmission maps to improve the modeling of the Agility MLC. Both models were implemented as prototypes in the RayStation TPS with the aim to match both the test fields typically used in the commissioning of the Agility MLC\cite{Snyder:2016aa,Roche:2018aa} and the methodology proposed by Saez et al\cite{Saez:2020aa}. Additionally, the study investigates the feasibility of using a common set of MLC parameters for different centers and assesses the impact of these new MLC models using measured clinically relevant plans.

\section{Materials and Methods}

Three different institutions with Elekta linacs (two VersaHD and one Synergy) equipped with the Agility MLC and the RayStation TPS participated in this study. Flattened beams with nominal energy 6\,MV were used at all centers.

\subsection{Tests and procedures}

\subsubsection{Sweeping gap and asynchronous sweeping gap tests}

Dynamic MLC sweeping gap (SG) and asynchronous sweeping gap (aSG) fields were used\cite{Saez:2020aa}. In SG tests, all leaves are placed at the same position, forming a rectangular aperture with a given gap size, and leaves move at a constant speed from left to right. The aSG are similar to SG, but adjacent leaves are offset a given distance $s$ in order to expose a controlled amount of the leaf edges, thus leading to a non-rectangular aperture shape (see Figure~\ref{fig:aSGsketch}) but the gap size between leaf pairs remains constant. The distance traveled by the MLC was set to 12\,cm and the dynamic sequence was defined with 13 control points. A dose rate of 500 MU/min was used and 200 monitor units (MU) were delivered for each field. The Y diaphragms were set to 10\,cm and gap sizes between leaf pairs of 5, 10, 20, and 30\,mm were used. The average doses for these tests were measured with a Farmer ion chamber at 10\,cm depth in water, and 90\,cm source-to-surface distance.

\newlength{\imagewidth}
\settowidth{\imagewidth}{\includegraphics{./figures/Figure1}}
\begin{figure*}[t!]
      \includegraphics[width = \imagewidth]{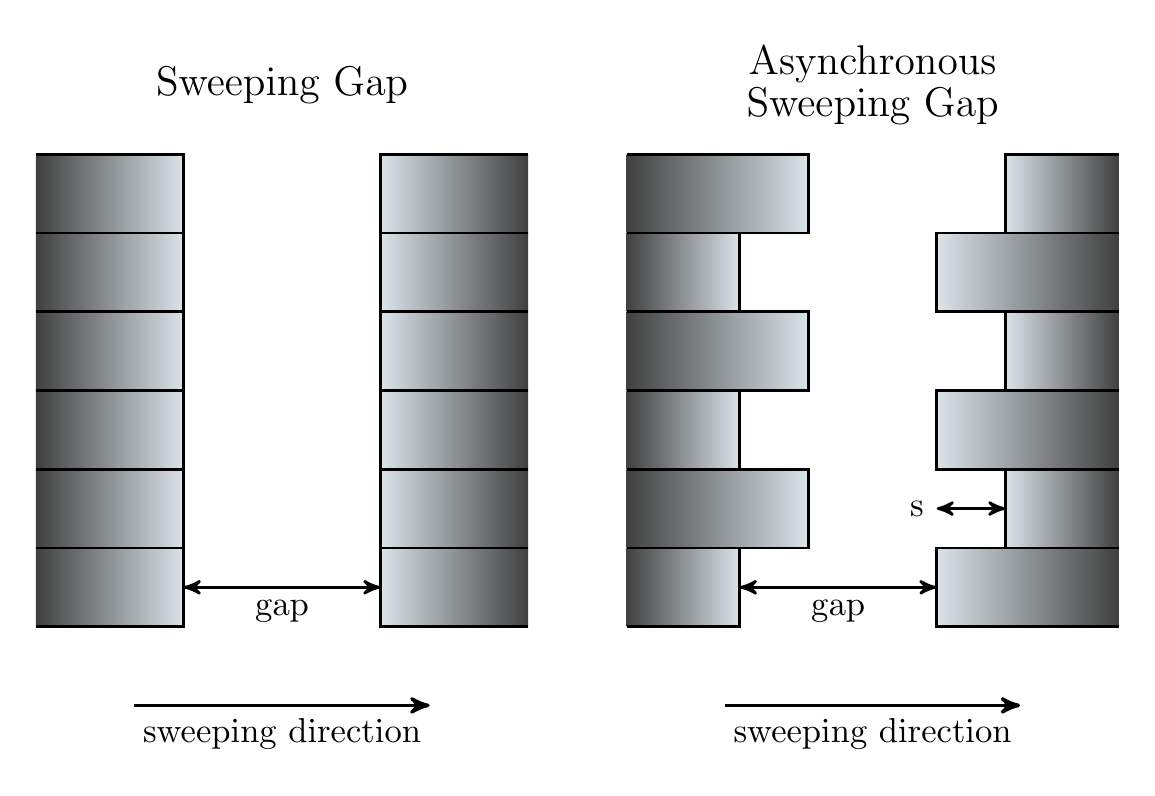}
	 \caption{Schematic of the sweeping gap and asynchronous sweeping gap tests. \vspace*{5mm}}
       \label{fig:aSGsketch}
\end{figure*}
\let\imagewidth\relax

The average MLC transmission was determined by measuring an open $10\times10$\,cm$^2$ field at 200\,MU, repeating with MLC bank A shielding the beam, then with MLC bank B shielding the beam. The average transmission is then simply calculated as the ratio of the average dose measured through both MLC banks to the open field dose.

The fluence reduction at the leaf side as a function of $s$, $\Delta \phi_{\text TG}$, is calculated following the methodology of Saez et al. using Equation~\ref{eq:deltaphi}\cite{Saez:2020aa}:

\begin{equation}
\Delta \phi_{\text TG} (s) = \frac{D_\text{SG} - D_\text{aSG}(s)}{2 k}
\label{eq:deltaphi}
\end{equation}

Where $k$ is a constant that relates fluence values to dose units which can be obtained from readings of the SG fields, the nominal leaf width at the isocenter plane (5\,mm for Agility MLC) and the average transmission\cite{Saez:2020aa}.

The same fields can be calculated in the TPS using a virtual water phantom with the dose scored to a cylindrical region-of-interest of 1\,cm diameter and 2\,cm length representing the collecting volume of the Farmer chamber. Thus, the $\Delta \phi_{\text TG}$  curves calculated by the TPS can be compared with experimental curves. 

\subsubsection{Static tests}

Elekta provides a set of predesigned tests, known as the ExpressQA package\cite{Snyder:2016aa,Roche:2018aa} as a tool to determine the MLC parameters in the TPS. The FOURL test was used to investigate tongue-and-groove modeling with three-channel film dosimetry performed using each center’s film protocol (home-made MATLAB software, radiochromic.com, and FilmQA Pro, respectively). 

\subsection{MLC models and parameter tuning}

\subsubsection{RayStation MLC transmission model (clinical model)}\label{sec:clinical}

The RayStation TPS uses an MLC model with constant transmission regions from which a fluence is computed using projection of virtual sources. The fluence is further traced into the patient using a collapsed cone or a Monte Carlo dose computation. For RayStation version 12A and earlier, the leaf tip and the tongue-and-groove regions are modeled with a constant transmission $\sqrt{T}$, where $T$ is the average MLC transmission. The widths of these regions can be configured with the parameters {\it leaf tip width} and {\it tongue-and-groove}. The tongue-and-groove parameter indicates the width of the affected regions sticking both outwards and inwards with respect to the nominal leaf width (all distances indicated at the isocenter plane). Thus, the total width of the tongue-and-groove region considered by the MLC model at each leaf side is twice the tongue-and-grove parameter. The leaf tip width parameter indicates the distance from the leaf tip end in which an increased transmission $\sqrt{T}$ is assigned. No tongue-and-groove at the leaf tip is considered in this MLC model; therefore, a  $\sqrt{T}$ transmission is assigned below the leaf tip and full transmission ($T=1$) is considered adjacent to the leaf tip. This MLC model is illustrated in Figure 2a and the transmission profiles below the leaves and below the outer tongue-and-groove region are shown in Figures 2d and 2e.

Three additional parameters ({\it offset}, {\it gain}, and {\it curvature}) can be used to define the difference between the leaf positions used in dose calculation and the DICOM/UI positions of the leaf tips. This RayStation transmission model gives a $\Delta \phi_{\text TG}$ which is zero until $s$ is equal to the leaf tip width and then increases linearly with $s$ with a slope which is proportional to the tongue-and-groove width\cite{Saez:2020aa}.

\begin{figure*}[t!]
      \includegraphics[width = \textwidth]{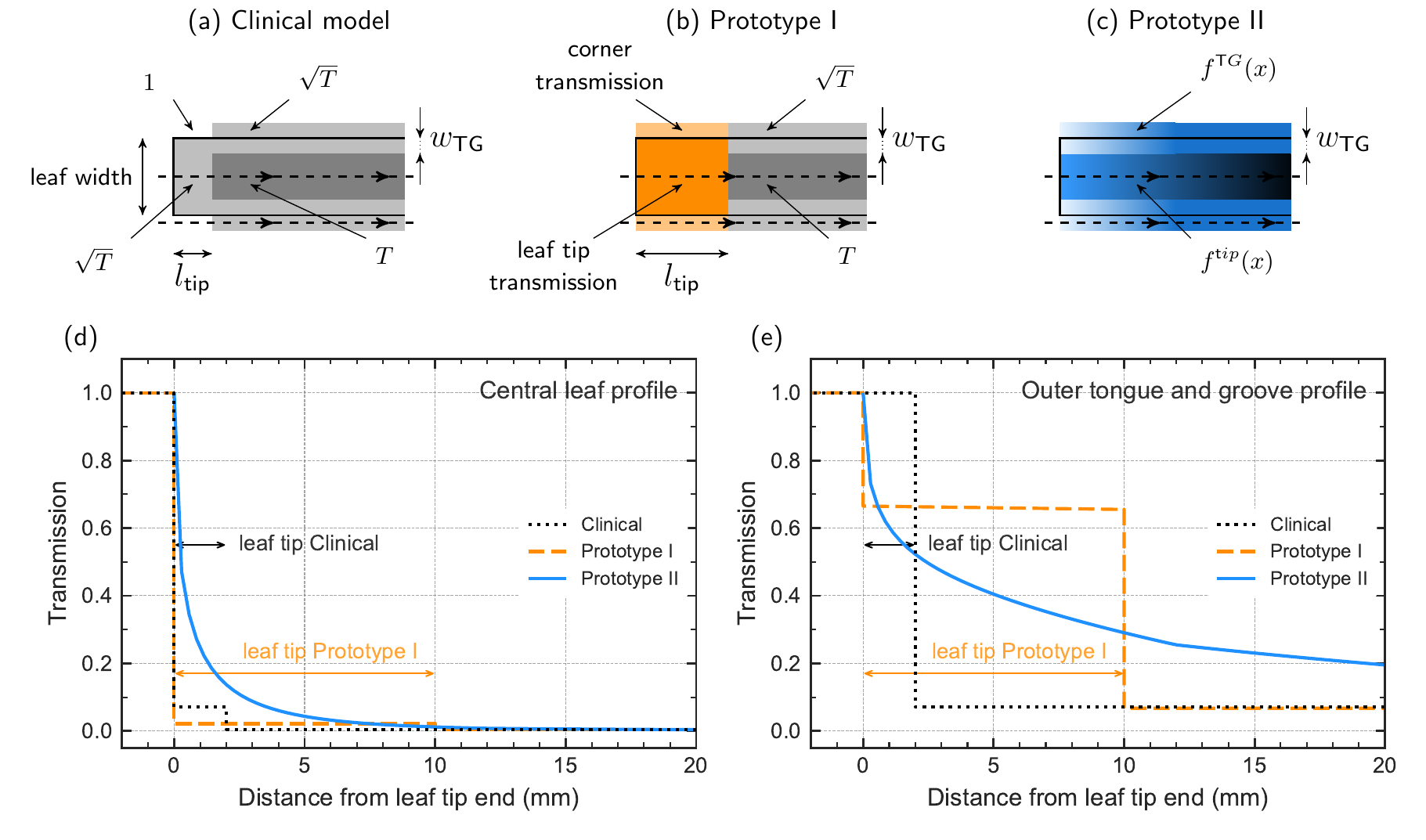}
	 \caption{(a) Schematic representation of the current clinical MLC model, (b) Prototype I, and (c) Prototype II. (d) Transmission profile under the center of the leaf width and (e) in the outer tongue-and-groove region as a function of the distance to the leaf tip end. The areas under different profiles cannot be directly compared due to the different tongue-and-groove widths used in each model. \vspace*{5mm}}
       \label{fig:models}
\end{figure*}

\subsubsection{Prototype I: Small changes to the RayStation MLC transmission model}

Prototype I is an evolution of the original MLC model in RayStation described in \ref{sec:clinical} where the transmission of the leaf tip and the transmission of the corner region (at the intersection between leaf tip and tongue-and-groove, where the leaf tip would have its tongue) were fitted to cross line profiles and to the measured $\Delta \phi_{\text TG}$. Adjusting the corner transmission to a value lower than 100\% produces a double linear fit in the calculated $\Delta \phi_{\text TG}$. curve instead of the constant plus linear fit described in Saez et al.\cite{Saez:2020aa}, allowing a better fit to the experimental $\Delta \phi_{\text TG}$ curve. A sketch of Prototype I is shown in Figure 2b, and the corresponding transmission profiles are given in Figures 2d and 2e. 

\subsubsection{Prototype II: Continuous MLC transmission model}

Prototype II models the variable transmission from the leaf tip towards the leaf base both below the leaf and over the tongue-and-groove region based on relative thickness of the known geometry and a realistic shadowing. The relative thickness and transmission of two adjacent leaves are combined into interleaf and interdigitation transmission functions. The transmission was given a lower bound of 0.45\%. The tongue-and-groove width away from the leaf tip was adjusted to fit the measured $\Delta \phi_{\text TG}$ curve and fixed to the same value for all centers. The MLC model of Prototype II is illustrated in Figure 2c and the continuous transmission profiles below the leaves and the tongue-and-groove regions are shown in Figures 2d and 2e.

\subsubsection{Configuration of the MLC models}

The RayStation MLC transmission models were used with the beam models and the MLC parameters clinically used at each center (clinical models), while both prototypes were used with a common set of MLC parameters at all centers except for the offset as explained below (see Table \ref{tab:parameters}). The common set of parameters were obtained by adjusting each prototype model to the average experimental $\Delta \phi_{\text TG}$  curve in order to obtain a good agreement with the measured synchronous and asynchronous sweeping gap doses. 

The only parameter that was tuned individually for each MLC is the leaf offset due to slight differences in the physical calibration of the MLC position on each machine. In center A, the offset was fitted by finding the best match between the Farmer-measured asynchronous 20\,mm gap test proposed by Saez et al.\cite{Saez:2020aa} and dose calculations. Center B tuned the offset based on optimizing the agreement between the measured central dose in Octavius4D and the calculated dose for five VMAT cases. In center C, the offset was tuned by optimizing the difference between the Farmer chamber dose in a low gradient region for 20 VMAT beams and the calculated dose. 

The implementation of the MLC parameters {\it offset}, {\it gain}, and {\it curvature} in the prototypes was not modified. However, the gain and curvature were set to zero in both Prototypes I and II and for all centers.

\subsection{Evaluation with clinical plans }

A set of clinically relevant plans were selected at each center, with the aim to cover a wide range of treatment sites and target sizes, and to select highly modulated plans in order to challenge the different MLC models. Supporting Figure S1 shows examples of different types of treatment plans included. Each center measured and evaluated computed dose with their clinical PSQA devices and software. 

\subsubsection{Center A}

Six different VMAT plans representing a wide range of geometrical and planning complexity were created. The plans had the characteristics outlined below.

A vertebra SBRT case with two different plans was generated. One was planned to achieve all the DVH criteria and maintain a ‘standard’ dose distribution. Another version was made to try to push the modulation as hard as possible. This was achieved by reducing the spinal cord objective to a very low value in the optimization resulting in a high MU (used as a surrogate for modulation). The intent was to introduce narrow leaf gaps throughout the arcs in order to test the sensitivity of the three calculation models to these narrow segments. Measurements for the two plans were made using EBT3 radiochromic film within a solid water phantom to investigate fine differences between the models. 

A dual-arc VMAT plan on a head and neck case and a single arc VMAT plan on a prostate patient dataset were included. These were measured using the PTW OctaviusII-729 detector array. 

The final two cases were made and measured directly in anthropomorphic phantoms. The first was a three-metastasis stereotactic radiosurgery (SRS) planned on the CIRS STEEV anthropomorphic phantom with a single isocenter and four coplanar arc plans. Measurements were made with EBT-XD radiochromic film in an axial plan and a sagittal plane. Three alanine pellets\cite{Sharpe:2000aa} were used to measure the dose inside the largest GTV and in the brainstem region. The second case was a spine SBRT plan made on a CIRS Atom phantom, in which the thorax-abdomen region was adapted in-house to allow for measurement of an axial EBT3 film and three alanine pellets through a region mimicking the T12 vertebra.

Dose calculations were made with 2\,mm dose grid spacing for the prostate and head and neck cases, and 1\,mm dose grid spacing for the SBRT and SRS cases. 

Analysis of radiochromic film and PTW Detector Array measurements was made using $\gamma$ index with settings: global $\gamma$ with dose difference relative to prescription dose, 20\% lower dose threshold, 2\%/2mm criteria. $\gamma$ passing rate as \% of points with $\gamma<1$ and mean $\gamma$ were recorded. An in-house software developed in MATLAB version 2018b - 2020b (The MathWorks Inc.) was used for the analysis.

\subsubsection{Center B}

A subset of fifteen plans was chosen out of a data set of 142 plans over a period of six months clinical routine according to the following criteria: all plans had challenging segments regarding tongue-and-groove exposure. Plans with different degrees of plan complexity were selected to cover a wide range of plan characteristics. All selected cases were VMAT. They were evaluated and judged to be clinically representative, covering paravertebral oligometastasis ($N=2$), brain metastastasis ($N=1$), prostate (N=3), para-aortic pelvic lymph nodes ($N=1$), eosophagus ($N=2$) and head and neck (oro- and hypopharynx, larynx, oral cavity) ($N=6$) plans. Dose calculations were made in RayStation with 2\,mm dose grid spacing.

The Octavius4D phantom along with the Octavius 1500 array were used for measurements. The PTW Verisoft v8.0 software was used to reconstruct a measured 3D dose and perform 3D gamma evaluation between reconstructed dose and TPS dose. A correction for daily output was applied. Gamma criteria were 2\%/2mm, of the maximum dose of the calculated volume and suppressing dose below 10\% of the maximum dose. Gamma calculation used 2nd and 3rd pass filter, helping to avoid false positive results caused by the resolution of the detector\cite{Depuydt:2002aa}. Gamma passing rate as percentage of points with $\gamma < 1$ were recorded.  

\subsubsection{Center C}

Ten different clinical VMAT plans covering different anatomical sites and levels of complexity were selected. Of these, five were lung SBRT plans, three with a single PTV and the two other plans with two PTVs. One of the lung plans used 360$^{\circ}$ VMAT dual-arcs whereas the other four plans used partial VMAT dual-arcs. Two plans were whole brain plans with hippocampal sparing and simultaneous multiple metastasis boost. The plans were highly modulated and optimized with three complete VMAT arcs. The last three plans were head and neck, treated to either two or three simultaneous dose levels.  These plans were optimized with 360$^{\circ}$ VMAT dual-arcs. Dose calculations were made with 2\,mm dose grid spacing.

All the plans were measured on the Sun Nuclear ArcCHECK diode array fixed on the table. Point dose measurements were also performed with a PTW N30001 Farmer chamber in a rectangular solid water phantom. 2-D analysis was performed using the Sun Nuclear SNC Patient software using the following $\gamma$ index settings: global $\gamma$ with absolute dose difference, 15\% lower dose threshold, 2\%/2mm criteria. The $\gamma$ passing rate was recorded as \% of points with $\gamma<1$.

\section{Results}

\subsection{Sweeping gap and asynchronous sweeping gap tests}

Figure~\ref{fig:doses} shows the measured SG and aSG tests with 40\,mm gap size, by the three centers, compared with the dose calculations from the different MLC models. For this gap size, the agreement between the measured and calculated by Prototype I and II are within 1\% across all $s$ values. However, the clinical models diverge from the measurements at $s > 10$\,mm for centers A and C, and $s > 20$\,mm for center B.

\begin{figure*}[b!]
\vspace*{5mm}
      \includegraphics[width = \textwidth]{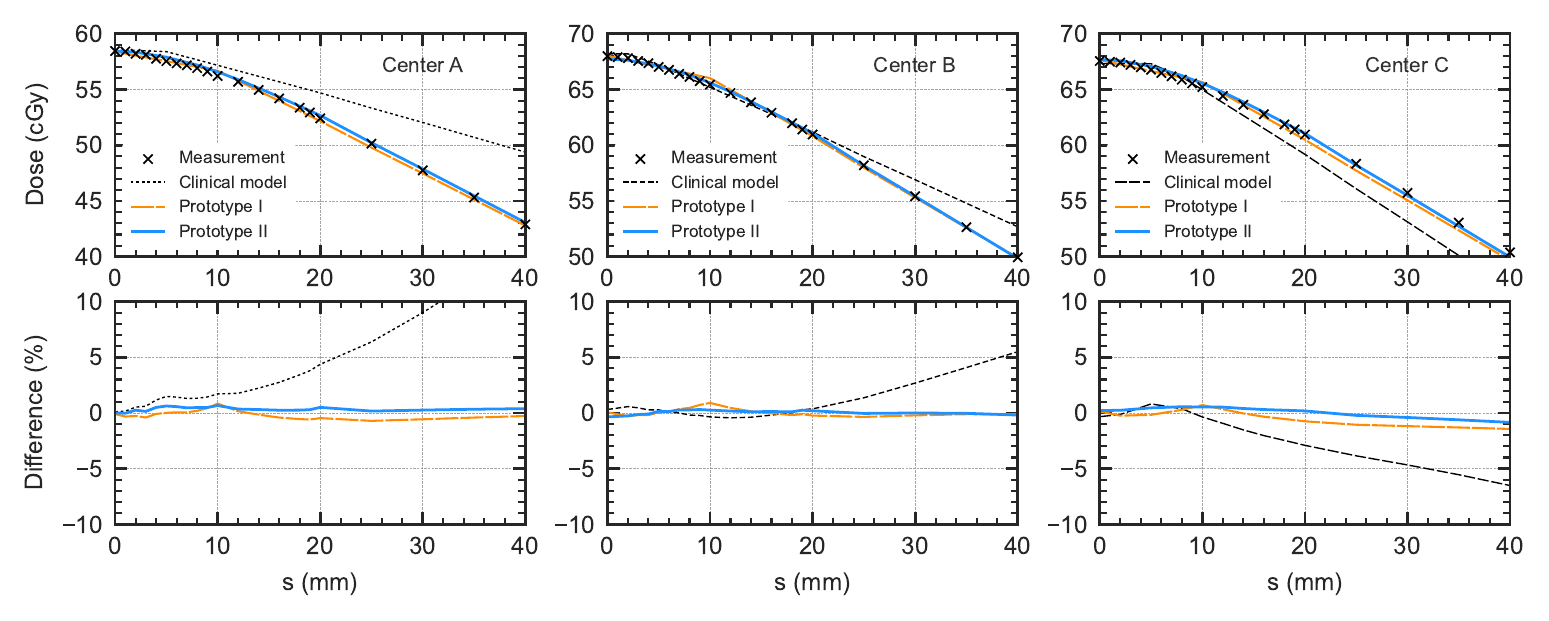}
	 \caption{Measured SG and aSG tests compared with the dose calculations from the different MLC models. The dose values are shown in the top graphs, and percentage difference given in the bottom graphs. The dose variation between the three centres is due to different monitor unit to dose calibration. \vspace*{5mm}}
       \label{fig:doses}
\end{figure*}
\begin{table}[h]
\small
\begin{center}
\begin{tabular}{llllll}
\hline\hline
Parameter 			&  Center A     	&   Center B 	&   Center C	& Prototype I	& Prototype II \\
\hline
Tongue-and-groove (cm)   	&       0.05		&    0.07		&   0.1		&  0.09		&  0.123         \\
Leaf tip width (cm)		&       0.46		&    0.25		&   0.6		&  1			&   -- \\
MLC transmission 		&       0.6\%	&    0.55\%	&   0.4\%		&  0.45\%		&  0.45\%\\
Offset (cm)          		&      -0.012 	&    -0.005		& -0.02		&  -0.006 to 0.015 & -0.077 to -0.093 \\
Gain          			&       0.0037	&    0.0007	&  0.008		&  0			&  0\\
Curvature (1/cm)          	&       0.0001	&    -0.000015	&  0			&  0			&  0\\
Leaf tip transmission & $\sqrt{T}$ & $\sqrt{T}$	&$\sqrt{T}$	&  2.15\% 		& Function of distance \\ 
& & & & & to leaf tip end\\
Tongue-and-groove transmission	& $\sqrt{T}$ & $\sqrt{T}$	&$\sqrt{T}$	&  $\sqrt{T}$	& Function of distance\\
& & & & & to leaf tip end\\
Corner transmission				&100\% 	&100\%& 100\% & 66.5\% 		& Function of distance\\
& & & & & to leaf tip end\\
\hline\hline
\end{tabular}
    \vspace*{-5mm}
    \caption{Summary with the MLC parameter values for the clinical, Prototype I and Prototype II MLC models. \vspace*{8mm}} \label{tab:parameters} 
\end{center}
\end{table}

\newlength{\imagewidth}
\settowidth{\imagewidth}{\includegraphics{./figures/Figure4}}
\begin{figure*}[b!]
\vspace*{5mm}
      \includegraphics[width = 1.5\imagewidth]{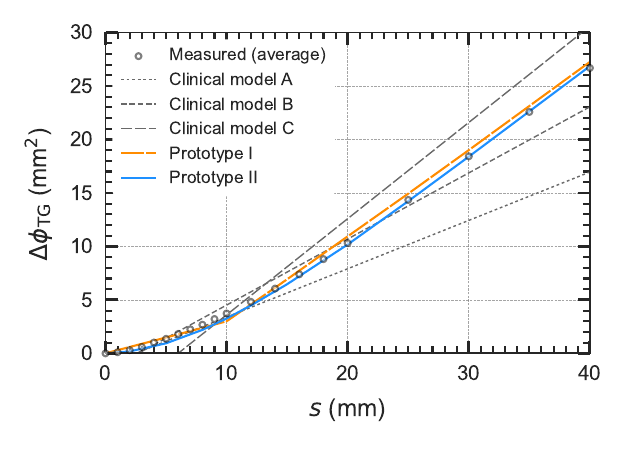}
	 \caption{The measured average  $\Delta \phi_{\text TG}$ curve from the three centers and computed values using the three clinical  models and the two prototypes. \vspace*{5mm}}
       \label{fig:deltaphis}
\end{figure*}
\let\imagewidth\relax

The $\Delta \phi_{\text TG}$ extracted from the aSG measurements represents the integrated transmission reduction by the tongue-and-groove along the leaf tip6. The measured $\Delta \phi_{\text TG}$ curves were consistent between centers but, as centers followed different commissioning strategies for their clinical MLC models, they arrived at different parameters in their clinical models (Table~\ref{tab:parameters}) and different computed $\Delta \phi_{\text TG}$ curves (Figure~\ref{fig:deltaphis}). The Center A beam model reproduced the $\Delta \phi_{\text TG}$ data for $s<12$\,mm, after which it started to underestimate the TG shadowing. The Center B beam model stayed relatively close to the measured curve all the way up to $s<25$\,mm but also underestimated the TG at larger distances from the leaf tip end. The Center C beam model had the same inclination as measurements at large $s$, with the cost of being far from the  $\Delta \phi_{\text TG}$ curve for most values of $s$. 

Both Prototype I and II were adjusted to fit the experimental $\Delta \phi_{\text TG}$ curve. Prototype I used three parameters to fit the  curve: the tongue-and-groove width, the leaf tip width and the transmission of the corner. For Prototype II the transmission functions were given by the geometry, and only the tongue-and-groove width was fitted to data. All these parameters were the same in all three centers (see Table~\ref{tab:parameters}). 

Figure~\ref{fig:deltaphis} shows how Prototype II follows experimental data almost perfectly for all values of $s$. The much simpler Prototype I can also be considered as a good fit.

\vspace*{-10mm}
\subsection{Static tests}

A profile from the FOURL film extracted at a region sensitive to the maximal tongue-and-groove shadowing and away from the leaf tip\cite{Hernandez:2022aa}. For a clinical RayStation model such a profile is best matched with a large tongue-and-groove width as used by Center C, whereas a clinical model obtained using the commissioning strategy of Center A underestimates both the depth of the valleys and the overall shadowing effect of the tongue-and-groove this far from the leaf tip (see Figure~\ref{fig:fourl} for center A, supporting Figure S2 gives the plots for centers B and C). Where compromises had to be made for the clinical model, Prototype I and II can follow the $\Delta \phi_{\text TG}$ and still be considered a good fit to the FOURL profile considering computation resolution (Figure~\ref{fig:fourl}).


\begin{figure*}[b!]
\vspace*{-10mm}
      \includegraphics[width = \textwidth]{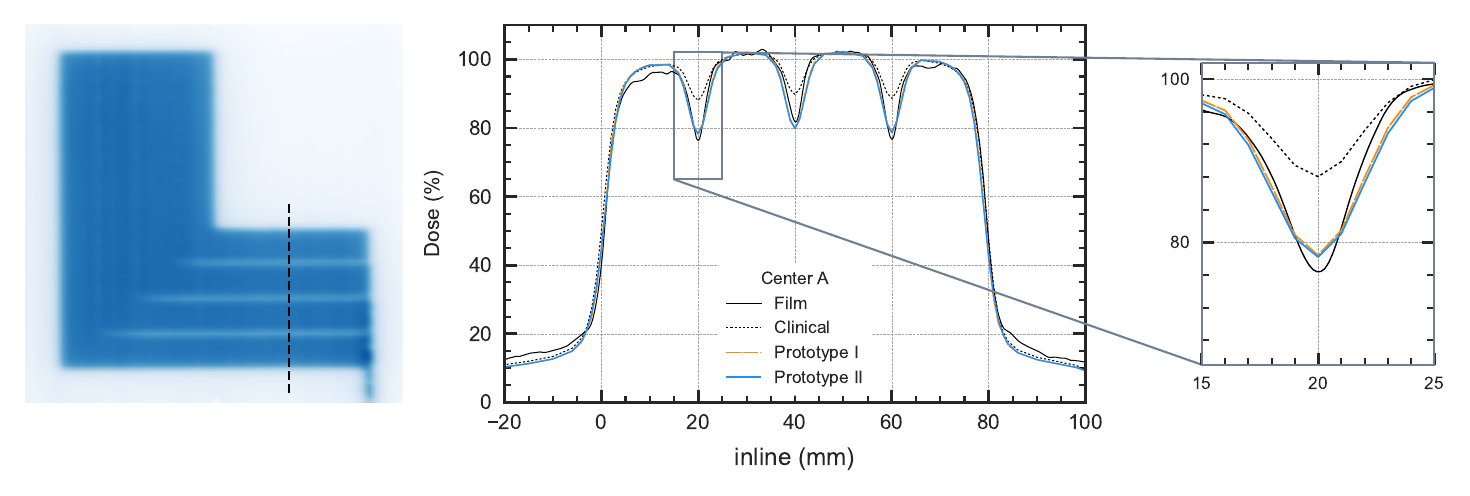}
	 \caption{FOURL test results. A dose map measured with film dosimetry for a representative center (center A) is shown in (a). Measured and calculated line dose profiles along the dashed line indicated in (a) are shown in (b). \vspace*{5mm}}
       \label{fig:fourl}
\end{figure*}

\subsection{Evaluation with clinical plans}

Figure~\ref{fig:psqa} shows a summary of all of the gamma passing rates, according to each center’s chosen analysis approach, and shows the difference between the passing rate for the clinical model and for Prototype I and II respectively. Example gamma index maps are shown in supporting information Figures S3 and S4.

In center A and C, there was no significant difference between the clinical model and the two prototype models for the plans measured with various equipment. The average gamma passing rates (GPR) at 2\%/2mm were $98.4\%\pm 2.0\%$ (clinical model), $98.7\%\pm 1.5\%$ (Prototype I), and $98.5\%\pm 2.0\%$ (Prototype II) for center A. For center C it was $96.9\%$ (clinical model), $96.9\%$ (Prototype I), and $96.4\%$ (Prototype II). The mean gamma value averaged over all the cases was similar for the clinical model and the two prototypes at $0.34 \pm 0.05$ and $0.44 \pm 0.1$ for center A and C respectively.  In the anthropomorphic plans of center A, the alanine measurements in the high dose regions were $-0.34\%$ (clinical), $0.37\%$ (Prototype I) and $1.15\%$ (Prototype II) for the SRS case, and $0.77\%$ (clinical), $0.65\%$ (Prototype I) and $0.94\%$ (Prototype II) for the spine SBRT case. The uncertainty on the alanine measurement is $1.7\%$ at one standard deviation. For center C, the average dose difference for Farmer measurements over the 10 cases were $-0.5\%\pm 1.8\%$ (clinical model), $-0.1\%\pm 1.3\%$ (Prototype I), and $0.2\%\pm 1.4\%$ (Prototype II).

\begin{figure*}[b!]
\vspace*{10mm}
      \includegraphics[width = \textwidth]{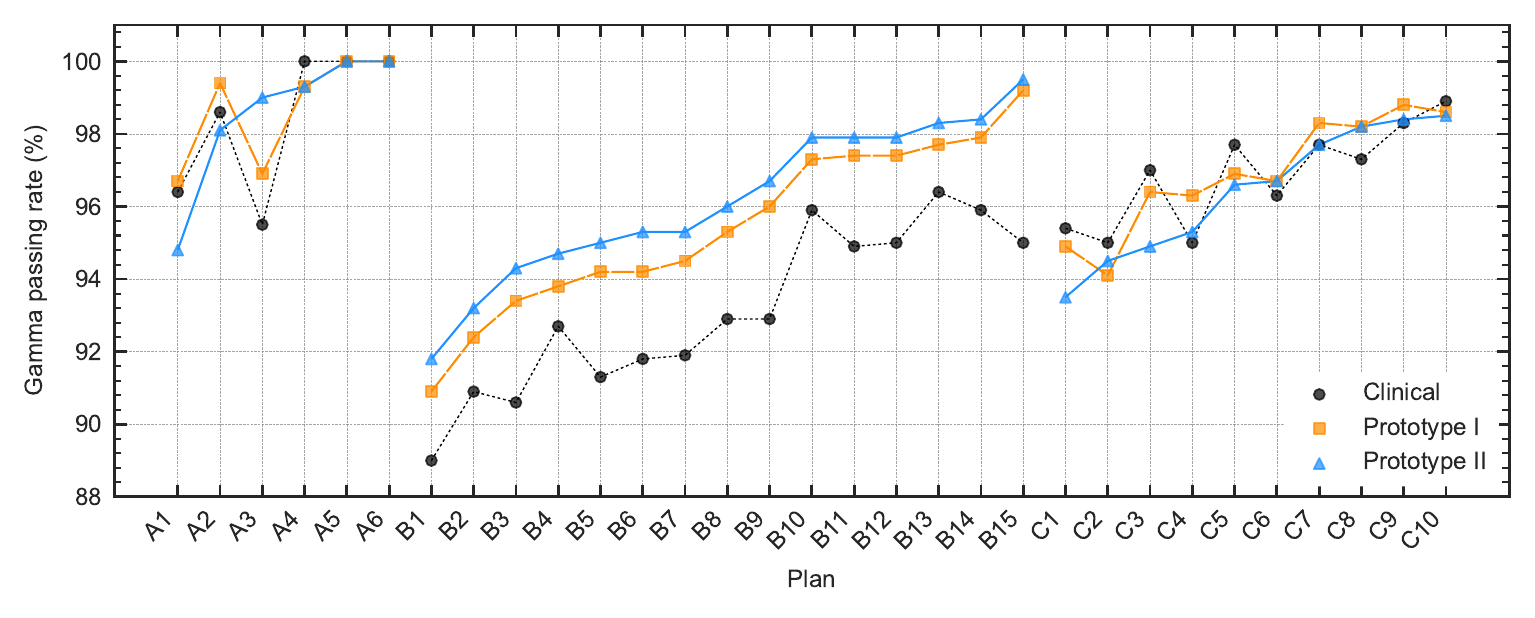}
	 \caption{Summary all of the PSQA gamma passing rates according to each center’s analysis for 2\%/2mm. The labels indicate center and plan number. \vspace*{5mm}}
       \label{fig:psqa}
\end{figure*}

At center B, the offset adjustment strategy was to minimize the average dose difference between computation and measurement at the phantom center using the PSQA-device itself. Plan 1, 3, 6, 11 and 13 was used for the tuning which also increased the gamma passing rate for the other plans.  Both Prototype I and Prototype II improved the agreement in clinical plans, with an increase in GPR of $2.3\%\pm 0.8\%$ and $3.0\pm 0.8\%$ , respectively. The average GPRs were $93.1\%\pm 2.3\%$ (clinical model), $95.4\%\pm 2.4\%$ (Prototype I), and $96.1\%\pm 2.2\%$ (Prototype II).

\section{Discussion}

\vspace*{-5mm}
The combination of rounded leaf tip and tilted leaf side of the Agility MLC leads to a complex transmission pattern around an extended leaf tip presenting challenges which have been shown to not be properly modelled in RayStation and Monaco\cite{Hernandez:2022aa}. This leads to variability in the MLC configuration in these treatment planning systems, resulting in large variations in the MLC parameters used by the community. The present study has shown that it is possible to improve the modelling of these characteristics and to facilitate easier MLC configuration in the TPS.

The asynchronous sweeping gap test\cite{Saez:2020aa,Hernandez:2017aa} characterized the integrated shadowing effect of the tongue-and-groove along the leaf tip and showed that the constant plus linear shape of the clinical RayStation $\Delta \phi_{\text TG}$  curve (versions 12A and earlier) could not replicate the sigmoid shape of the Agility $\Delta \phi_{\text TG}$ curve over the full range of distances to the leaf tip end ($s = 0$ to 40\,mm).  Without this detailed mapping, a clinic must decide which region of the sigmoid curve to focus on, giving a model that may be accurate for some types of plans but possibly less so for others, generally underestimating the dose reduction from the tongue-and-groove far from the leaf tip. On the other hand, fitting MLC parameters to static test fields, such as the FOURL profile from the Elekta ExpressQA-package\cite{Snyder:2016aa,Roche:2018aa}, results in a large tongue-and-groove width that does not provide a good fit close to the leaf tip end.

Two new MLC transmission models were developed within a prototype version of RayStation 12B and evaluated using the same tests. Prototype I, which was only a slight modification of the current MLC model in RayStation, was able to fit the  $\Delta \phi_{\text TG}$  curve and the FOURL y-profile with a minimum amount of added complexity, assigning a constant transmission at the leaf tip and at the intersection between the leaf tip and the tongue-and-groove. This prototype provided a better agreement with the experimental aSG doses and the $\Delta \phi_{\text TG}$  curve. 

Prototype II, which starts from first principles using a variable transmission along the leaf, could follow the curved shape of the $\Delta \phi_{\text TG}$  curve and the depth of the FOURL $y$-profile, but fitted it only slightly better than Prototype I. 

The small variation of the $\Delta \phi_{\text TG}$  curve between centers indicates that it is sufficient to tune MLC transmission parameters to an average behavior within each MLC class. Only the offset was adjusted for each machine and prototype, while the other MLC parameters were kept constant.  

The three competing transmission models were compared for clinical plans. In some cases, a slight improvement was obtained with Prototype II, but the accuracy of both prototypes was quite similar. In general, both prototypes provided an accuracy at least as good as the clinical models used at each center. At centers A and C, the differences between prototypes and clinical models were minor, while the prototypes outperformed their clinical model at center B (Figure~\ref{fig:psqa}). This could be because at center B the offset was adjusted based on PSQA results, which could be compensating for slight limitations in the IMRT QA device. Such a  strategy might jeopardize the independence of IMRT QA devices and result in TPS configuration errors in case systematic errors are present in such devices. We believe that adjusting the offset parameter based on ion chamber measurements of sweeping gap tests and clinical plans is preferable and also allows the use of IMRT QA devices as verification systems that are truly independent of the TPS configuration.

The main advantage of both prototypes is that they fit the full range of the $\Delta \phi_{\text TG}$  curve without requiring any compromises. The clinical models evaluated, on the other hand, require compromises and can accurately fit only part of the $\Delta \phi_{\text TG}$  curve, which will likely maximise dosimetric accuracy for plans with characteristics within a given range and be less optimal for others. This is in agreement with previous publications reporting that the optimal MLC parameters depend on the plan characteristics, which can be explained by the existing limitations in MLC models \cite{Hernandez:2022aa, Kielar:2012aa, Middlebrook:2017aa, Vieillevigne:2019aa}. For example, large target volumes treated with VMAT may require a good overall compromise over the full $\Delta \phi_{\text TG}$  curve, whereas small targets treated with IMRT/VMAT typically involve small MLC apertures that may require an accurate fit for small $s$. Such compromises and limitations could potentially be reduced by using the proposed MLC model improvements and commissioning strategy. Consequently, MLC prototypes presented in this study may be valid over a wider range of treatment plans, as tested by the three centers and shown in Figure~\ref{fig:psqa}, resulting in more robust MLC models. 

This extended validity was also supported by the results obtained with the FOURL test, which is commonly used in the commissioning of the Agility MLC. Thus, even if the characteristics of this static test are far from clinical plans, a good agreement was obtained with the prototypes at all centers. 

Another important advantage of the prototypes is that a single set of MLC parameters could be used at all centers, with only a minor adjustment of the offset (within $\pm 0.02$\,cm) to compensate for small differences in the specific calibration of the MLC position on each linac. This greatly facilitates the center's MLC configuration process, which is a critical step in modern techniques but can be challenging and cumbersome. It also allows the establishment of clear references that are useful to reduce the variability in the parameters used by the community and to improve safety in clinical practice\cite{Glenn:2022aa,Glenn:2020aa,Kerns:2017aa}. 

The fact that the two prototypes performed comparably for both commissioning test fields and clinical plans indicates that a highly sophisticated MLC model considering all the fine spatial details is not necessarily superior to a simpler MLC model when both models have been properly designed and fine-tuned to the asynchronous sweeping gap data. Hence, simple MLC models can provide accurate calculations over a wide range of clinical plan characteristics and be easily configured using sweeping gap fields. More sophisticated MLC models, for instance based on detailed raytracing or Monte Carlo calculations, could be beneficial for validation and benchmarking even if this study indicates they are not strictly needed for clinical dosimetry or the selected test fields.

One limitation of this study is that only the Agility MLC and the 6 MV energy was evaluated. We focused on the Agility because its modeling was shown to be particularly challenging and in need for improvement\cite{Hernandez:2022aa}. A good agreement was already reported for Varian’s Millennium and HD120 MLCs, but the same MLC prototypes could also be applied to these and other types of MLCs. It would be beneficial in particular for those MLCs with tilted tongue-and-groove like the MLCi and Unity MLC\cite{Saez:2020aa,Hernandez:2017aa}. Regarding the energy, it was already shown that the optimal MLC parameters for a given MLC type are independent of the nominal energy\cite{Saez:2020aa}, but the presented procedure can be readily applied to other energies.

\newpage
\section{Conclusions}

The aSG/SG tests provide a robust tool for MLC evaluation, comparison, and commissioning, and they expose limitations in MLC models and guide their improvement. The two MLC prototypes created within RayStation for the Agility accurately replicated the measured SG/aSG doses, facilitating the standardization of the configuration and commissioning processes and extending their range of validity. The similarity of the results between centers indicates that the same set of MLC parameters can be used, reducing the workload required for this task and its associated risks , and allowing standardization. All MLC models achieved good accuracy in clinical plans, but the prototypes improved the agreement in some cases, notably in the FOURL test. The simple prototype performed similarly to the more sophisticated one both in tests and clinical plans and constitutes a good option for routine clinical practice. Additionally, this simple MLC model reduced the need for trade-offs and was successfully configured using a common set of MLC parameters across different centers, which can be useful for reducing the workload and risks associated with the MLC configuration process and for improving the accuracy and safety of radiotherapy treatments.

\section{Acknowledgments}

This work was supported by the National Measurement System of the UK’s Department for Business, Energy and Industrial Strategy.

\section{Conflict of Interest Statement}

NPL authors report institutional collaboration agreement with RaySearch Laboratories. Agnes Angerud is an employee of RaySearch Laboratories. The rest of the authors have nothing to disclose.

\newpage
\section*{References}
\addcontentsline{toc}{section}{\numberline{}References}
\vspace*{-35mm}






\vspace*{20mm}
\bibliography{./agility_prototypes}      



\bibliographystyle{./medphy}    

\section*{Supporting information}
Additional information is available online in the Supporting Information section.

%
\let\imagewidth\relax
\let\imageheight\relax
\end{flushleft}
\end{document}